\documentclass[aps,prc,twocolumn,showpacs,floatfix,nofootinbib]{revtex4} 
\usepackage{graphicx}
\usepackage{dcolumn}
\usepackage{amsmath}
\usepackage{mathptmx}
\usepackage{booktabs}
\begin{document}

\title{Radiative strength functions in $^{163,164}$Dy}

\author{H.~T.~Nyhus$^1$\footnote{Electronic address: h.t.nyhus@fys.uio.no}, S.~Siem$^1$, M.~Guttormsen$^1$, 
A.~C.~Larsen$^1$, A.~B\"{u}rger$^1$, N.~U.~H.~Syed$^1$, G.~M.~Tveten$^1$ and A.~Voinov$^2$}

\affiliation{$^1$Department of Physics, University of Oslo, N-0316 Oslo, Norway}
\affiliation{$^2$Department of Physics and Astronomy, Ohio University, Athens, OH 45701, USA}

\date{\today}

\begin{abstract}
The nuclei $^{163,164}$Dy have been investigated using the Oslo method on data from the pick-up reaction $^{164}\rm{Dy(^{3}He,\alpha\gamma)^{163}Dy}$ and the inelastic scattering $^{164}\rm{Dy(^3He,^3He'\gamma)^{164}Dy}$, respectively. The radiative strength functions for both nuclei have been extracted, and a small resonance centered around $E_{\gamma} \approx 3$ MeV is observed in both cases. The parameters of this so-called pygmy $M1$ resonance (the scissors mode) are compared to previous results on $^{160,161,162}{\rm Dy}$ using the Oslo method, and to data on $^{163}{\rm Dy}$ measured by the Prague group using the two-step cascade method. In particular, the integrated reduced transition probability $B(M1\uparrow)$ of the pygmy resonance is compared with neighboring dysprosium isotopes. We also observe an enhanced strength in the region above $E_{\gamma} \approx 5$~MeV in $^{164}\rm{Dy}$. Possible origins of this feature are discussed.
\end{abstract}  

\maketitle

\section{Introduction}
A continuing effort has long been devoted to study $\gamma$ decay from excited nuclei. 
The radiative strength function (RSF) represents the mean value of the decay probability via a $\gamma$ ray with 
energy $E_{\gamma}$, and contains rich information on the average electromagnetic properties of the nucleus. 

For high-energy $\gamma$ transitions ($\sim 7-20$~MeV), the RSF is dominated by the giant electric dipole resonance (GEDR). 
At lower energies other resonances have been discovered, such as the giant magnetic dipole resonance (GMDR, also called the giant magnetic spin-flip resonance) and the electric quadrupole resonance; however, these have a significantly lower strength~\cite{RIPL2}. In addition, there are other structures observed in the RSF governed by various collective modes of the nucleus. These structures are often referred to as pygmy resonances due to their low strength compared to the GEDR. There are two known pygmy resonances: the $E1$ resonance for $\gamma$ ray energies between $5-10$~MeV that is believed to stem from neutron skin oscillation~\cite{skin}, and the $M1$ resonance called the scissors mode, which is observed in the region of $E_{\gamma} = 3$~MeV for rare-earth nuclei~\cite{scissor}. 

The RSFs below the neutron threshold have been studied mainly by ($\gamma,\gamma'$) experiments, also called nuclear resonance fluorescence (NRF)~\cite{NRF}. 
Other methods, such as the two-step-cascade (TSC) method~\cite{TSC} and the Oslo method~\cite{Sch00a}, have also successfully provided data on the RSF for many nuclei. The latter method enables us to extract the RSF for $\gamma$ ray energies up to the neutron binding energy $B_n$. This method has been used in the present analysis.\\

Previously, experiments have been performed on $^{160,161,162}{\rm Dy}$~\cite{bagheri} using the Oslo method~\cite{Sch00a}. From these data, the widths of the $M1$ pygmy 
resonance have been found to be about two times greater than the width found for $^{163}{\rm Dy}$ obtained by the Prague group using the TSC method~\cite{tscp}. In the present work we have studied $^{163,164}{\rm Dy}$ to investigate the discrepancy between the measured widths. In particular, we have compared the total integrated strength $B(M1\uparrow)$  of all the mentioned Dy isotopes.

Details about the experimental method are presented in Section II, followed by the experimental results for the RSF in Section III. Finally, conclusions will be drawn in Section IV. 

\section{Experimental procedure and data analysis} 
\label{sec:to}
\begin{table*}
\caption{Parameters used for normalizing $\rho$ and $\cal T$.} 
\begin{tabular}{lccccccccccc}
\hline
\hline
	
Nucleus    & $E_{\mathrm{pair}}$ & $C_1$ & $a$                  & $D$      &$\sigma(B_n)$& $B_n$ &$\rho(B_n)$  &  $J_t$&$\left<\Gamma_{\gamma} \right>$ & $\eta $\\ 
                   & (MeV)  		             & (MeV)   &  (MeV$^{-1}$) & (eV)   &(MeV)   &(MeV)             &  ($10^{6}$ MeV$^{-1}$)&  & (meV) & \\
\hline

$^{163}$Dy  & 0.661 & $-1.293$ & 17.653 & 62(5)   &  5.492&  6.271   & 0.96(12)   & 0  & 112 & 0.52 \\
$^{164}$Dy  & 1.707 & $-1.291$ & 17.747 & 6.8(6)& 5.408 & 7.658 & 1.74(21)   &$\frac{5}{2}$ & 113 &0.56    \\
\hline
\hline
\end{tabular}
\\
\label{tab:GogC}
\end{table*}
The experiment was conducted at the Oslo Cyclotron Laboratory (OCL), using a $38$-MeV beam of $^3{\rm He}$ particles. The target of $98.5\%$ enriched $^{164}{\rm Dy}$ had a thickness of $1.73$~mg/cm$^2$, and the reactions $\rm{^{164}Dy(^{3}He,\alpha\gamma)^{163}Dy}$ and  $\rm{^{164}Dy(^3He,^3He'\gamma)^{164}Dy}$ were studied.

The $\gamma$ rays and ejectiles were measured with the CACTUS multidetector array~\cite{CACTUS}, which consists of a sphere of 28 collimated NaI $\gamma$ detectors with total efficiency of $15\%$ of $4\pi$, surrounding a vacuum chamber which contains eight $\Delta E-E$ Si particle telescopes with thicknesses of 140 and 1500~$\mu{\rm m}$, respectively. The particle telescopes were placed in forward direction at $45^{\circ}$ relative to the beam axis.

\begin{figure}
\includegraphics[width=8.6cm]{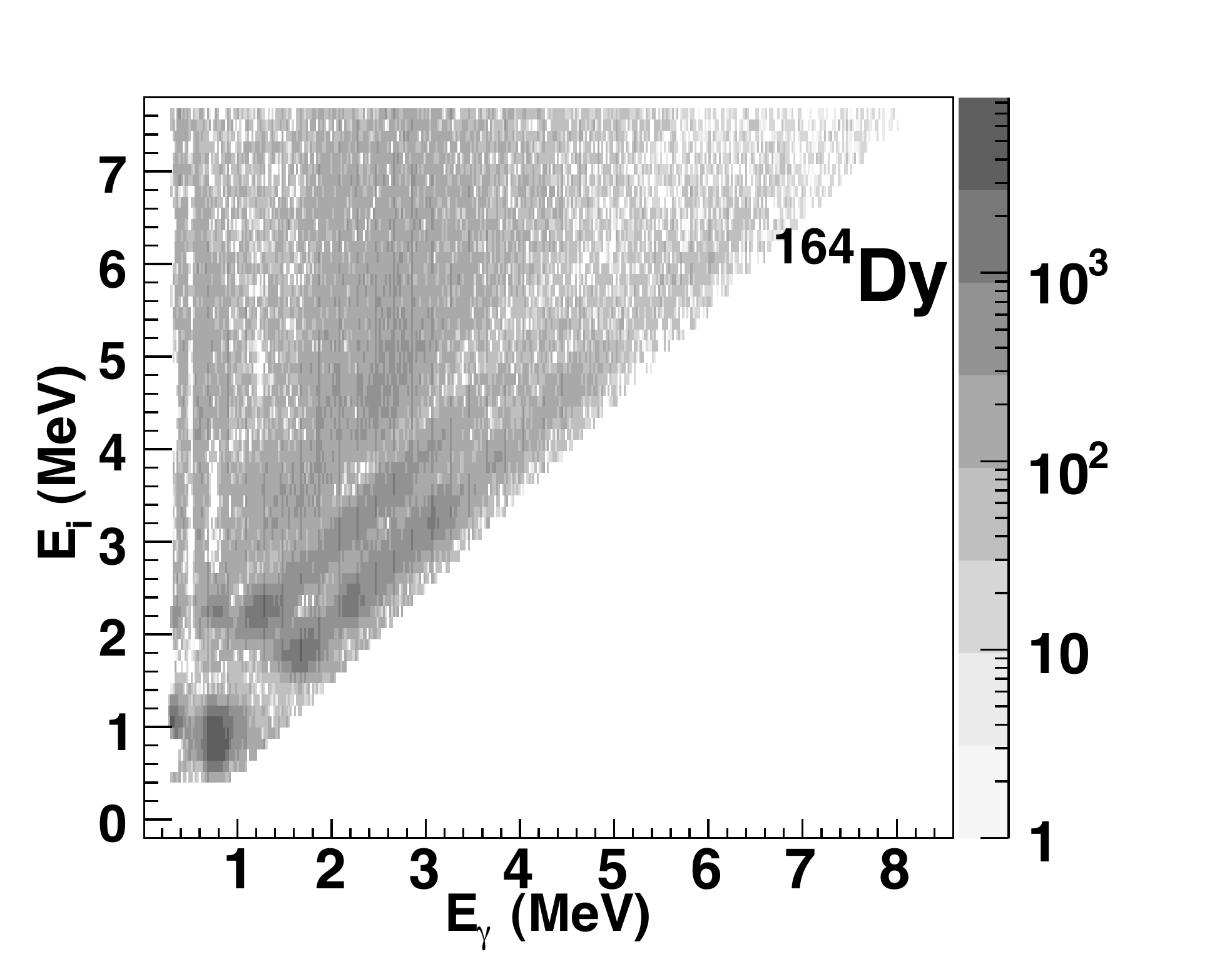}
\caption{The primary $\gamma$ ray matrix for $^{164}{\rm Dy}$, displaying primary $\gamma$ rays emitted at each initial excitation energy.}
\label{fg}
\end{figure}

From the known $Q$-values the excitation energy of the nuclei were calculated from the detected ejectile energy using reaction kinematics. The particles and 
$\gamma$ rays were measured in coincidence; hence, each $\gamma$ ray can be assigned to an initial excitation energy of the nucleus. The $\gamma$ ray spectra 
were unfolded using the known response functions of the CACTUS detector array~\cite{unfold}. 
The excited nuclei will decay through a cascade of $\gamma$ rays 
down to the ground state. By using the first-generation method~\cite{fg}, we were able to isolate the first (primary) $\gamma$ rays emitted in each $\gamma$ decay cascade. The distribution of primary $\gamma$ rays was found for each excitation energy bin, giving an excitation energy vs. $\gamma$ ray energy matrix denoted by $P(E_i,E_{\gamma})$. The primary $\gamma$ ray spectrum was normalized to unity for each excitation energy bin, which then represents the decay probability for each $\gamma$ ray with energy $E_{\gamma}$ decaying from a certain excitation energy $E_i$: $\sum_{E_{\gamma}=E_{\gamma}^{\mathrm{min}}}^{E_i} P(E_i,E_{\gamma})=1$. 
The primary $\gamma$ ray matrix for $^{164}{\rm Dy}$ is shown in Fig.~\ref{fg}. The diagonal line of the matrix corresponds to decay directly to the ground state ($E_{\gamma}=E_i$); however, it is more probable to decay to the first excited $2^+$ and $4^+$ states. Therefore, a ridge is formed in the matrix for decay to $E_i \sim 150$~keV, see Fig.~\ref{fg}.

The original Brink-Axel hypothesis~\cite{Brink, Axel} states that the GEDR can be built on every excited state, and that the properties of the GEDR 
do not depend on the temperature of the nuclear state on which it is built. This hypothesis can be generalized to include any type of collective excitation. 
Provided that this hypothesis is valid,  the primary $\gamma$ ray matrix can be factorized as
\begin{equation}
P(E_i,E_{\gamma}) \propto {\cal T}(E_{\gamma})\rho(E_i-E_{\gamma}),
\label{BrinkAxel}
\end{equation}
where $P(E_i,E_{\gamma})$ is the experimentally obtained and normalized primary $\gamma$ ray matrix. The function ${\cal T}(E_{\gamma})$ represents the 
radiative transmission coefficient and $\rho(E_i-E_{\gamma})$ is the level density at the final energy $E_f=E_i-E_{\gamma}$. 
The above factorization is based on the 
essential assumption that the system is fully thermalized prior to $\gamma$ emission, so that the reaction can be described as a two-stage process of which the first 
is the formation of the compound nucleus, which subsequently decays in a manner that is independent of the mode of formation~\cite{BohrMottelson}. 
The formation of a complete compound state is as fast as $\sim10^{-18}$~s, significantly less than the typical life time of a state in the quasi-continuum 
which is $\sim10^{-15}$~s. Therefore, the assumption is believed to be reasonable, and the decay process 
is at least mainly statistical. Recently, it has been shown that Eq.~(\ref{BrinkAxel}) can be valid even in some cases where full thermalization is not achieved~\cite{bly}. 

However, there is experimental evidence that the Brink-Axel hypothesis is violated for high temperatures (above 1-2 MeV). In particular, the width of the GEDR has been 
shown to depend on the temperature of the final states~\cite{PL383}. For our experimental conditions the excitation energy (and thus the temperature) 
is relatively low and changes slowly with excitation energy ($T\sim \sqrt{E_f}$). Therefore, we assume that the radiative strength function does not 
depend on temperature in the energy region under consideration.

The right hand side of Eq.~(\ref{BrinkAxel}) is normalized to unity, yielding
\begin{equation}
P(E_i,E_{\gamma}) = \frac{ {\cal T}(E_{\gamma})\rho (E_i-E_{\gamma})}{\sum_{E_{\gamma} '=E_{\gamma}^{min}} {\cal T}(E_{\gamma}')\rho (E_i-E_{\gamma}')}.  
\label{nr3}
\end{equation}
Using this equation and applying a least squares fit to the primary $\gamma$ ray matrix, a unique functional form of $\rho(E_i-E_{\gamma})$ and 
${\cal T}(E_{\gamma})$ is derived~\cite{Sch00a}, while the normalization is yet to be determined. There are infinitely many normalization options that 
reproduce the experimental primary $\gamma$ ray matrix. All the solutions are related to each other through the two transformations~\cite{Sch00a}
\begin{equation}
\widetilde{\rho}(E_{i}-E_{\gamma}) = A\:\exp[\alpha(E_{i}-E_{\gamma})]\:\rho(E_{i}-E_{\gamma})
\label{eq:6c}
\end{equation}
and
\begin{equation}
\widetilde{{\cal T}}(E_{\gamma}) = B\: \exp(\alpha\,E_{\gamma})\:{\cal T}(E_{\gamma}),
\label{eq:6d}
\end{equation} 
where $A, B$ and $\alpha$ are constants representing the absolute value of $\rho(E_i-E_{\gamma})$ and ${\cal T}(E_{\gamma})$, and the slope of the two functions, respectively.  These parameters are determined by normalizing Eqs.~(\ref{eq:6c}) and~(\ref{eq:6d}) to known experimental data. The parameters $A$ and $\alpha$ are identified by normalizing the experimental level density to known levels found from discrete spectroscopy at low energies. 
\begin{figure}[htb]
\includegraphics[width=8.6cm]{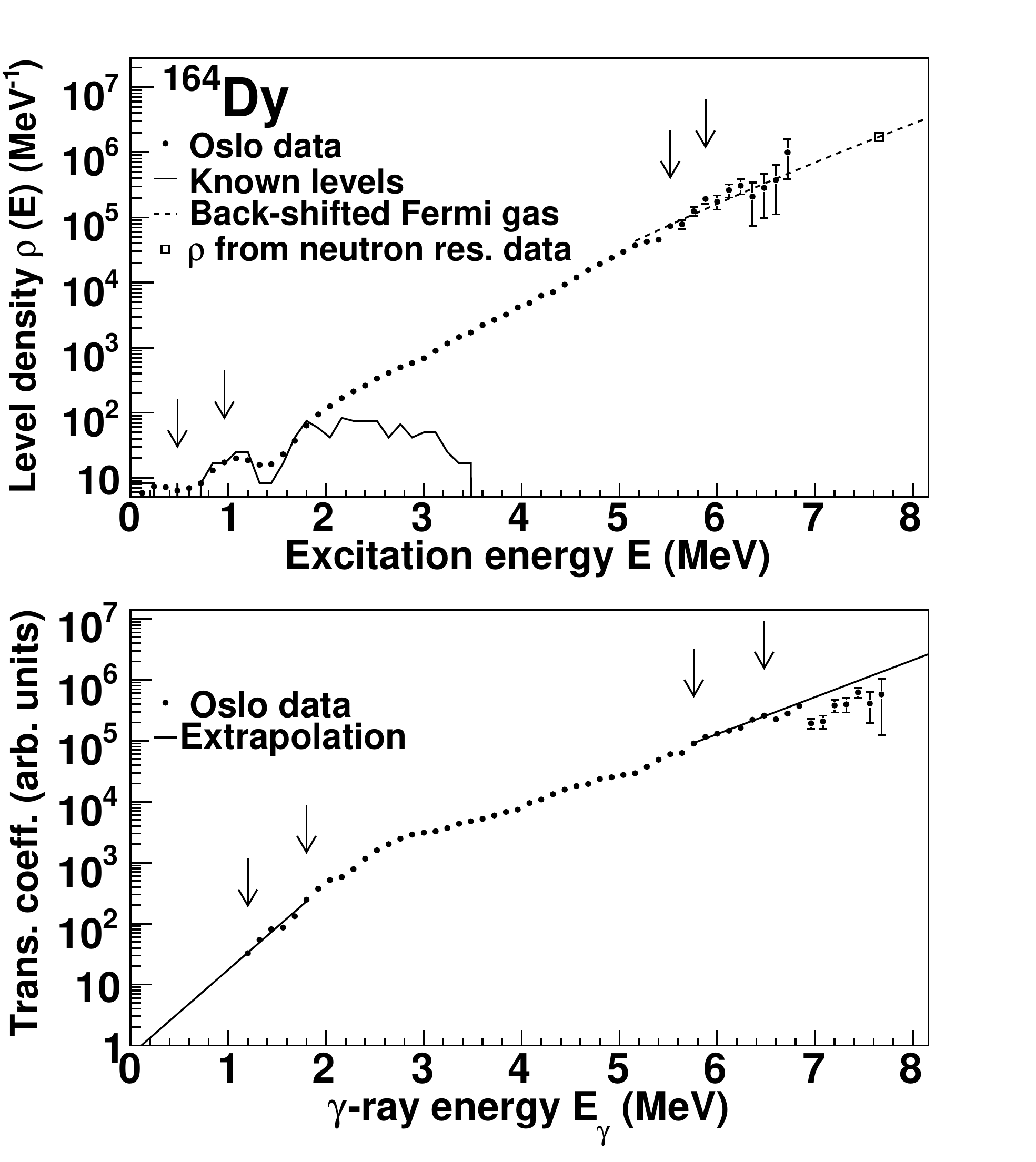}
\caption{Upper panel: The level density of $^{164}\rm{Dy}$, normalized  to known discrete levels and to $\rho(B_n)$ calculated from neutron resonance spacing data, with an interpolation using the back-shifted Fermi gas model. Lower panel: the radiative transmission coefficient of $^{164}\rm{Dy}$ with extrapolations. The normalization is performed in the regions between the arrows.}
\label{fig:rhotrans}
\end{figure}
At higher excitation energies the experimental level density is normalized to the level density determined from the known neutron resonance spacing data~\cite{RIPL2} at the neutron binding energy $B_n$. 
The present experimental data extend up to about $B_n-1$~MeV; an interpolation is thus required to reach $B_n$. The back-shifted Fermi gas model~\cite{Gil65,esb} was applied for this purpose:
\begin{equation}
\rho_{bs}(E)=\eta \frac{\exp(2\sqrt{aU})}{12\sqrt{2}a^{1/4}U^{5/4}\sigma},
\label{eq:RhoBS}
\end{equation}
where the constant $\eta$ is applied to adjust $\rho_{bs}(E)$ to the semi-experimental level density at $B_n$. The intrinsic excitation energy is given by $U=E-C_1-E_{{\rm pair}}$, where $C_1$ is the back-shift parameter equal to $C_1=-6.6A^{-0.32}$~MeV, where $A$ represents the mass number. The pairing energy $E_{{\rm pair}}$ is based on the pairing gap parameters $\Delta_p$ and $\Delta_n$ evaluated from odd-even mass differences~\cite{Wapstra} according to~\cite{bohrmottel}. The parameter $a=0.21A^{0.87}$~MeV$^{-1}$ corresponds to the level density parameter. The spin-cutoff parameter $\sigma$ is given by $\sigma^2=0.0888aTA^{2/3}$, where the nuclear temperature is described by 
\begin{equation}
T=\sqrt{U/a}.
\end{equation}
The normalization of $\rho(E_i-E_{\gamma})$ for $^{164}\rm{Dy}$ is displayed in the upper panel of Fig.~\ref{fig:rhotrans}.

Finally, ${\cal T}(E_{\gamma})$ is normalized by determining the coefficient $B$, which gives the magnitude of ${\cal T}(E_{\gamma})$. We have the following relation between the total radiative width of neutron resonances $\left< \Gamma_{\gamma}\right>$ at the neutron binding energy and the radiative transmission coefficient ${\cal T}(E_{\gamma})$~\cite{kop}:
\begin{equation}
\left< \Gamma_{\gamma}\right>=\frac{1}{4\pi\rho(B_n,J^{\pi}_i)}\sum_{J^{\pi}_f}\int_0^{B_n}dE_{\gamma}B{\cal T}(E_{\gamma})\rho(B_n-E_{\gamma},J^{\pi}_f),
\label{eq:int}
\end{equation}
where $D_i=1/\rho(B_n,J^{\pi}_f)$ is the average spacing of $s$-wave neutron resonances. The summation and integration extends over all final levels with spin $J_f$ which are accessible by $\gamma$ radiation with energy $E_{\gamma}$. 
 
Due to methodological difficulties, ${\cal T}(E_{\gamma})$ cannot be determined experimentally for low-energy $\gamma$ rays, $E_{\gamma}~<~1$~MeV~\cite{trans}. 
In addition, the data suffer from poor statistics for $\gamma$ ray energies $E_{\gamma}>B_n-1$~MeV. We therefore extrapolate ${\cal T}(E_{\gamma})$ with an exponential function, as demonstrated for $^{164}\rm{Dy}$ in the lower panel of Fig.~\ref{fig:rhotrans}. For further details of the normalization procedure, see Ref.~\cite{trans}. The parameters used for normalizing $\rho(E_i-E_{\gamma})$ and ${\cal T}(E_{\gamma})$ are given in Table~\ref{tab:GogC}.

Note that the uncertainties displayed in Fig.~\ref{fig:rhotrans} only reflect statistical uncertainties, and do not include the uncertainties related to the model used for normalization. This is also the case for the other figures showing experimental data.

\begin{table*}[!htb]
\caption{Parameters used for the radiative strength functions.} 
\begin{tabular}{lccccccccccc}
\hline
\hline
	
Nucleus    & $E_{E1}^1$ & $\sigma_{E1}^1$ & $\Gamma_{E1}^1$     & $E_{E1}^2$ & $\sigma_{E1}^2$ &$\Gamma_{E1}^2$  & $E_{M1}$ & $\sigma_{M1}$ & $\Gamma_{M1}$ &  $\beta_2$\\ 
           & (MeV)  		 & (mb)  &  (MeV) & (MeV) & (mb)      & (MeV)  & (MeV)  & (mb) & (MeV) &     \\
\hline

$^{163}$Dy  &12.37 &278.50   & 3.17  &15.90   &139.04  & 5.12  &7.51  &1.49  & 4.00  & 0.300   \\
$^{164}$Dy &12.26 &280.41   & 3.12  &15.95   &140.00  & 5.15  &7.49  &1.49  & 4.00  & 0.314   \\
\hline
\hline
\end{tabular}
\\
\label{tab:RIPL2}
\end{table*}

\begin{table*}[!htb]
\caption{Fitted pygmy-resonance parameters and normalization constants.} 
\begin{tabular}{lccccc}
\hline
\hline
	
Nucleus    & $E_{py}$ & $\sigma_{py}$ & $\Gamma_{py}$ & $\kappa$ \\ 
          & (MeV) & (mb) & (MeV)&    \\
\hline

$^{163}$Dy  &2.81(9) &0.72(12)   & 0.86(19)  & 1.78(14)     \\
$^{164}$Dy &2.81(6) &0.53(6)   & 0.80(12)  &1.72(6)    \\
\hline
\hline
\end{tabular}
\\
\label{tab:fit}
\end{table*}

\section{Radiative strength functions} 
\begin{figure}
\includegraphics[width=8.6cm]{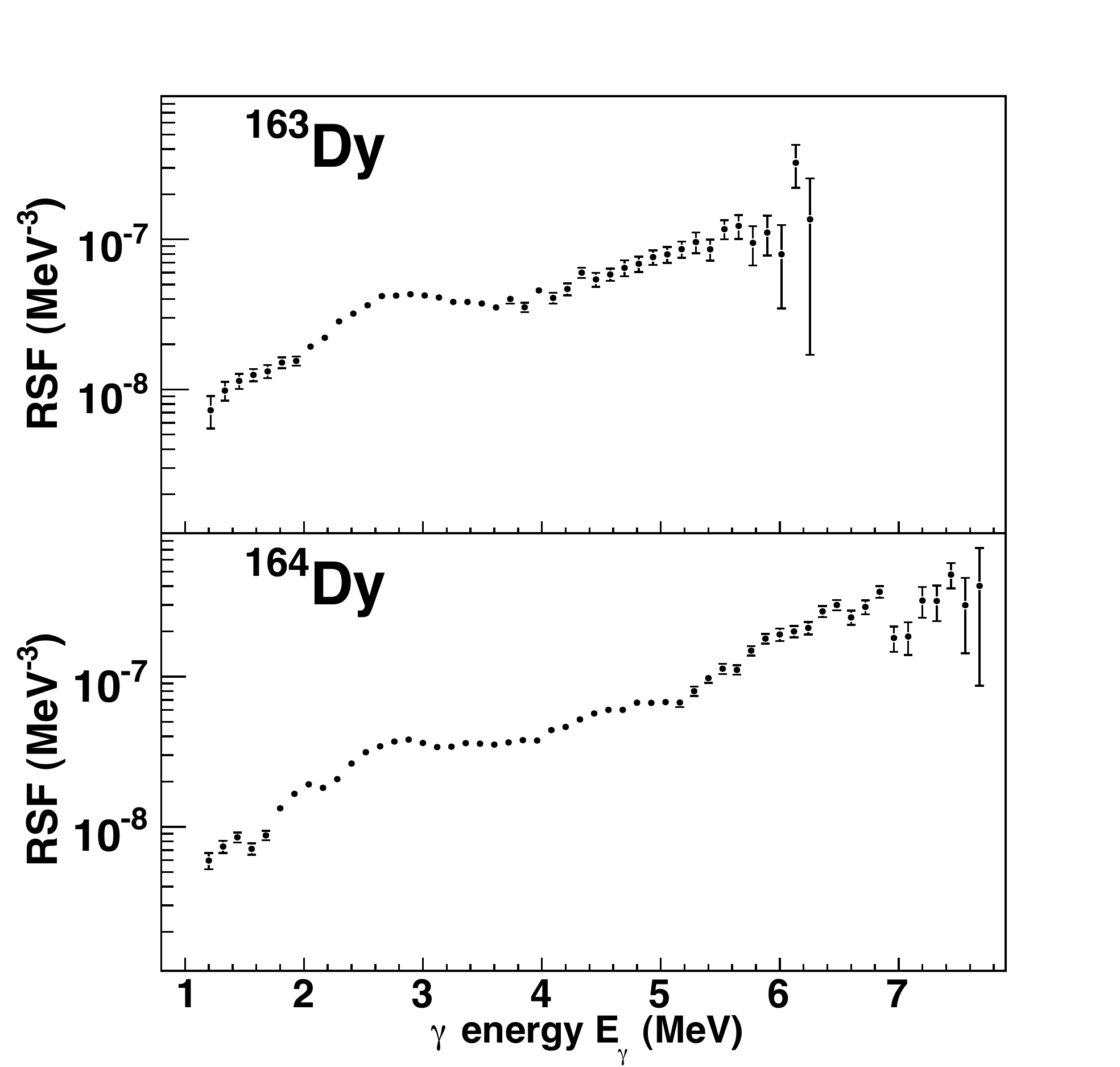}
\caption{Normalized RSFs of $^{163,164}{\rm Dy}$.}
\label{fig:rfs}
\end{figure}

Assuming that $\gamma$ decay taking place in the quasi-continuum is dominated by dipole transitions ($L=1$), the radiative strength function can be calculated from the normalized transmission coefficient by
\begin{equation}
f(E_{\gamma})=\frac{1}{2\pi}\frac{{\cal T}(E_{\gamma})}{E^3_{\gamma}}.
\end{equation}
Using the above relation, we obtain the experimental RSFs displayed in Fig.~\ref{fig:rfs}. We observe that they are increasing functions of $\gamma$ energy, and we can easily identify the $M1$ pygmy resonance in both cases. 
We expect the RFS to be composed of the pygmy resonance, the giant electric dipole resonance (GEDR) and the giant magnetic dipole resonance (GMDR). The Kadmenski\u{i}-Markushev-Furman (KMF) model~\cite{Kad83} is employed to characterize the $E1$ strength. In this model, an excitation-energy dependence is introduced through the temperature of the final states $T_f$:
\begin{equation}
f_{E1}^{\rm{KMF}}(E_{\gamma})=\frac{1}{3\pi^2\hbar^2c^2}\frac{0.7\sigma_{E1}E_{\gamma}\Gamma^2_{E1}(E_{\gamma}^2+4\pi^2T_f^2)}{E_{E1}(E^2_{\gamma}-E_{E1}^2)^2},
\label{eq:gedr}
\end{equation}
where $\sigma_{E1}$, $\Gamma_{E1}$ and $E_{E1}$ denote the peak cross section, width and the centroid of the GEDR, respectively. In general, the KMF model describes experimental data very well; however, the temperature dependence violates the Brink-Axel hypothesis. In line with the previously mentioned argument that the temperature varies relatively little in our region of interest, we have assumed that the temperature can be considered to be 
constant. Thus, the Brink-Axel hypothesis is revived.

It was found that a constant temperature of $T_f=0.3$~MeV gives a good fit to the experimental data, in agreement with Ref.~\cite{bagheri}. For deformed nuclei the 
GEDR is split into two, and is therefore described as a sum of two strength functions given by Eq.~(\ref{eq:gedr}). The GMDR is thought to be governed 
by the spin-flip $M1$ resonance~\cite{trans}, and can be described by a Lorentzian function:
\begin{equation}
f_{M1}(E_{\gamma})=\frac{1}{3\pi^2\hbar^2c^2}\frac{\sigma_{M1}E_{\gamma}\Gamma_{M1}^2}{(E_{\gamma}^2-E_{M1}^2)^2+E_{\gamma}^2\Gamma_{M1}^2},
\label{GMDR}
\end{equation}
where $\sigma_{M1}$, $\Gamma_{M1}$ and $E_{M1}$ give the peak cross section, width and the centroid of the GMDR, respectively. The GEDR and GMDR parameters are taken from the systematics of Ref.~\cite{RIPL2} calculated with the deformation parameter $\beta_2$~\cite{RIPL2}. The $M1$ pygmy resonance $f_{py}$ is described by a Lorentzian function similar to the one given in Eq.~(\ref{GMDR}). All parameters are listed in Table II.

The theoretical strength function is then given by
\begin{equation}
f=\kappa(f_{E1}+f_{M1})+f_{py},
\label{eq:rsfth}
\end{equation}
where $f_{E1}$, $f_{M1}$ and $f_{py}$ represent the contributions from the GEDR, GMDR and the $M1$ pygmy resonance, respectively. The parameter $\kappa$ is a normalization constant. Together with the pygmy resonance parameters $\sigma_{py}$, $\Gamma_{py}$ and $E_{py}$, $\kappa$ is used as a free parameter when performing a least squares fit to adjust the total theoretical strength to the experimental data. \\

\begin{figure}[htb]
\includegraphics[width=8.6cm]{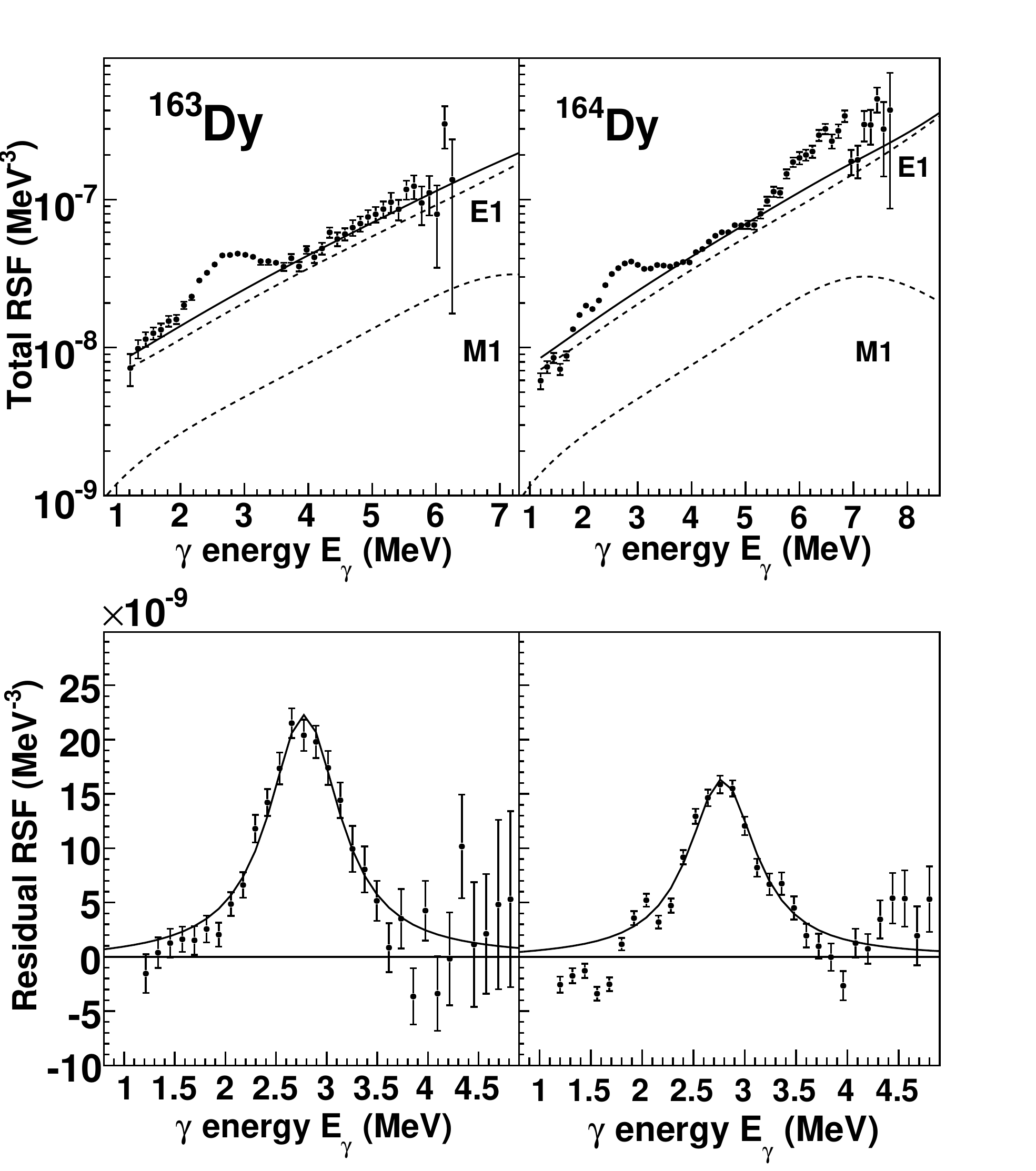}
\caption{Experimental RSFs of $^{163}\rm{Dy}$ (left panel) and $^{164}\rm{Dy}$ (right panel). The dashed lines in the uppermost panels show the contributions from the giant dipole resonances multiplied by $\kappa$, the solid line represents the sum of these two contributions, $\kappa(f_{E1}+f_{M1})$. The fit to the experimental datapoints in $^{164}{\rm Dy}$ is performed up to $E_{\gamma} = 5.3$~MeV. In the lower panels the sum $\kappa(f_{E1}+f_{M1})$ is subtracted from the experimental data and the fit to the $M1$ pygmy resonance is displayed (solid line).}
\label{fig:fitRSF}
\end{figure}

The fit to the experimental data points is shown in Fig.~\ref{fig:fitRSF} for both nuclei. The upper panels show the contributions $\kappa f_{E1}$ and $\kappa f_{M1}$ and the sum of these two contributions. 
In the lower panels the sum $\kappa(f_{E1}+f_{M1})$ is subtracted from the experimental data, and the fit to the $M1$ pygmy resonance is 
displayed. We notice that the fit to the experimental data around the $M1$ pygmy resonance is good, especially for $^{163}{\rm Dy}$. 
When comparing the pygmy resonance parameters of $^{163,164}{\rm Dy}$ (see Table~\ref{tab:fit}) to those extracted for $^{160,161,162}{\rm Dy}$ 
reported in~\cite{bagheri}, we find a smaller width of the pygmy resonance. The previous measurements for $^{160,161,162}{\rm Dy}$ yielded widths in the range of $\Gamma_{py}=1.26-1.57$~MeV 
using a constant temperature of $T_f=0.3$~MeV. In the present work, with the same constant temperature, we find widths of $\Gamma_{py}=0.86$~MeV and $\Gamma_{py}=0.80$~MeV for $^{163}{\rm Dy}$ 
and $^{164}{\rm Dy}$, respectively. The nucleus $^{163}{\rm Dy}$ has been investigated earlier by the Prague group, analyzing TSC spectra from the 
$^{162}\textrm{Dy}(n, 2\gamma)^{163}\textrm{Dy}$ reaction~\cite{tscp}. In their work, the width of the pygmy resonance was reported to be $\Gamma_{py}=0.6$~MeV. For this specific case ($^{163}{\rm Dy}$), the measured $\Gamma_{py}$ from the Oslo data and the data from the Prague group is comparable within the uncertainties.\\

\begin{figure}
\includegraphics[width=8.6cm]{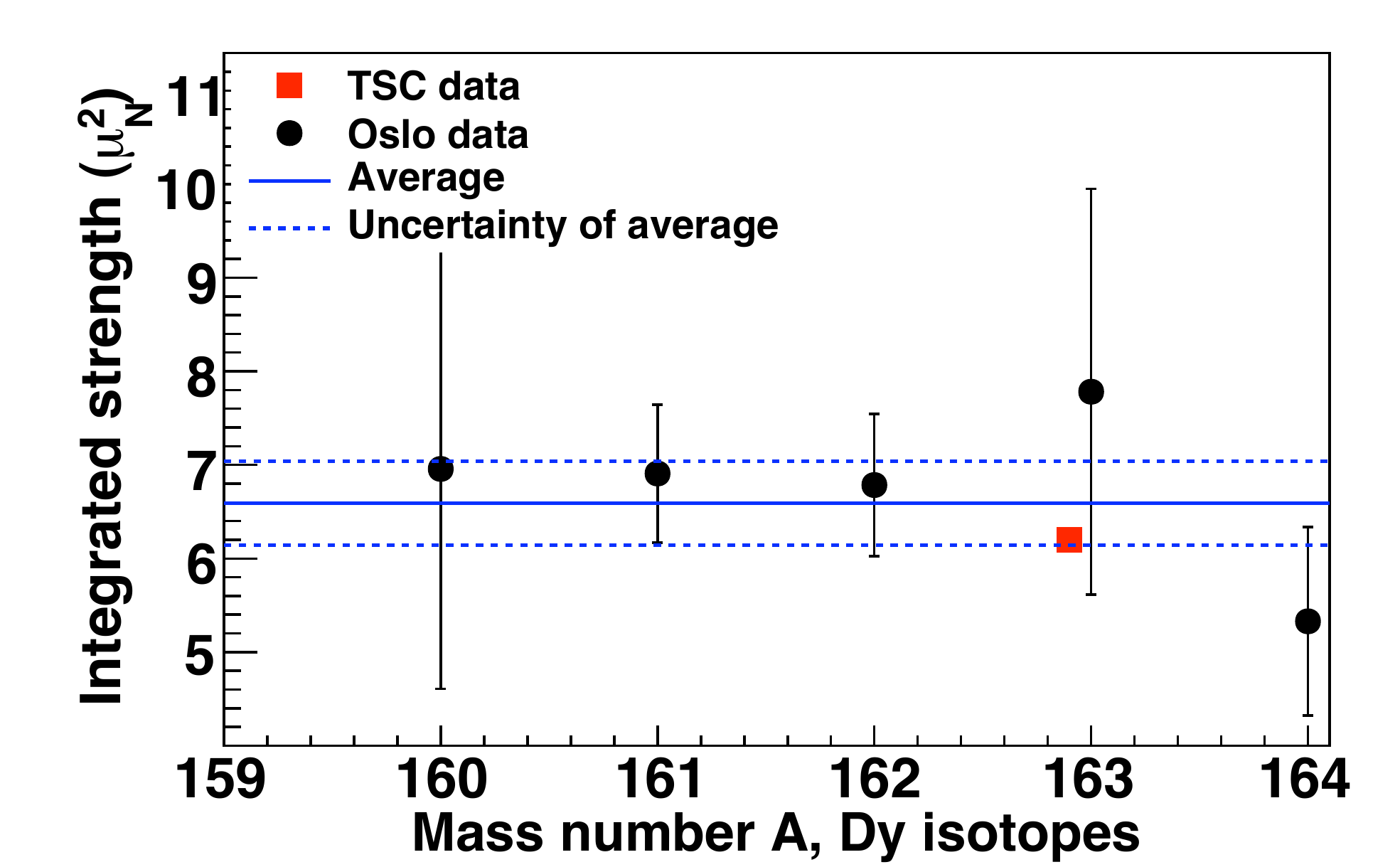}
\caption{[Color online] The integrated $B(M1~\uparrow)$ strength of the pygmy resonance for dysprosium isotopes measured with the Oslo method (filled circles) and their average value (solid line). The TSC data point for $^{163}{\rm Dy}$~\cite{tscp} is also displayed (filled square).}
\label{fig:B}
\end{figure}
We note from Fig.~\ref{fig:fitRSF} that $\sigma_{M1}$ for $^{163}{\rm Dy}$ is significantly larger than for $^{164}{\rm Dy}$. The reason for this is not yet understood. In order to obtain a more precise comparison, the total integrated strength $B(M1\uparrow)$
 given by
\begin{equation}
B(M1\uparrow)=\frac{9\hbar c}{32\pi^2}\left(\frac{\sigma \Gamma}{E}\right)_{M1 pygmy},
\end{equation}
is calculated for $^{160-164}{\rm Dy}$, and the results are displayed in Fig.~\ref{fig:B}. When calculating the weighted average of the Oslo data, a value of $6.6(4)$~$\mu_N^2$ is 
found\footnote{The $B(M1~\uparrow)$ values 
for $^{160,161,162}{\rm Dy}$ are calculated from Ref.~\cite{Sch00a}. The values of $^{161,162}{\rm Dy}$ are the weighted averages of the values obtained from 
the ${\rm(^{3}He,\alpha)}$ and ${\rm(^{3}He,^{3}He')}$ reactions.}. The $B(M1\uparrow)$ value from the TSC experiment is not included in the fit because no errors are given in Ref.~\cite{tscp}. We observe that all 
the measured values agree within the uncertainties.

\begin{figure}[htb]
\includegraphics[width=8.6cm]{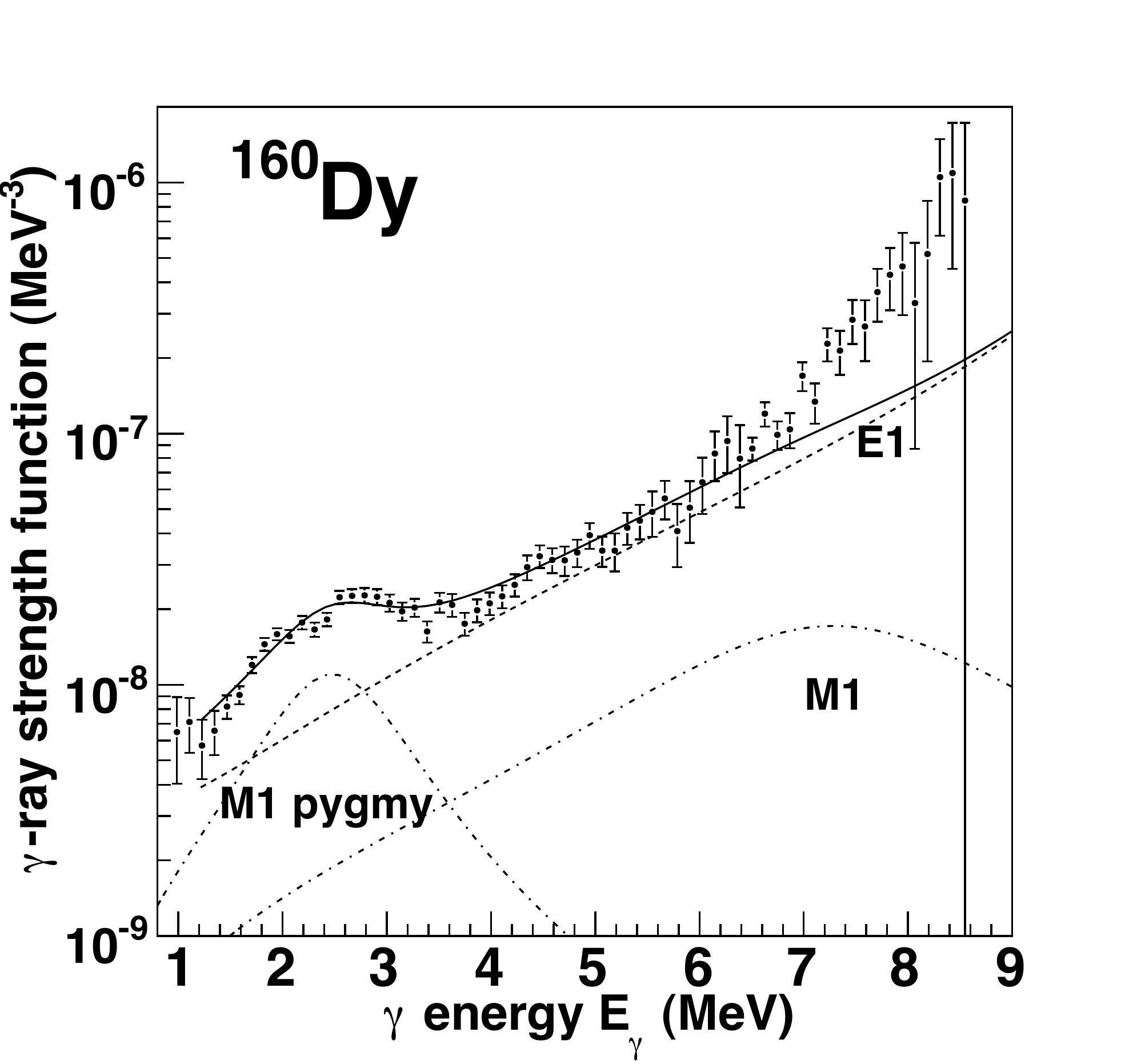}
\caption{Experimental RSF for $^{160}\rm{Dy}$. The dashed line represents the tail of the GEDR, while the dashed-dotted lines give the contributions from the GMDR and the $M1$ pygmy resonance. The solid line is the sum of all the resonances. The fit to the experimental datapoints is performed up to $E_{\gamma} = 6.9$~MeV.}
\label{fig:RSF160}
\end{figure}

For $^{164}{\rm Dy}$, we observe an increase in the RSF compared to theory 
for energies above $E_{\gamma}\approx 5.0$~MeV. Similar 
features have been observed in $(\gamma,\gamma')$ 
experiments on other nuclei, e.g., $^{116,124}{\rm Sn}$~\cite{PRC57} and $^{208}{\rm Pb}$~\cite{PRL89}. 
For these nuclei the structure is thought to be governed by the so-called neutron skin oscillation, 
a collective mode of $E1$ character that for stable nuclei is located in the region of 
$E_{\gamma}=5-10$ MeV. This feature has been observed in nuclei with a high neutron-to-proton ratio $N/Z$, 
and is interpreted as an 
oscillation of the neutron-enriched periphery of the nucleus versus a core consisting of equally 
many protons and neutrons, $N=Z$~\cite{skin,C50}. Enhanced strength is also observed in the 
RSF of $^{117}\rm Sn$ measured at OCL~\cite{UAC}. Unfortunately, the present experimental technique 
cannot provide information on the electromagnetic character of the enhanced strength in $^{164}\rm{Dy}$. 
However, it might be a reasonable guess that the observed strength stems from the $E1$ skin oscillation, 
since we note that Dy nuclei have a high neutron-to-proton ratio of 
$N/Z=1.36-1.48$ for the stable isotopes. Evidence of both the $M1$ pygmy resonance and the $E1$ pygmy resonance in one and the same nucleus has, however, 
not been reported earlier. 

Data on $^{160}\rm{Dy}$ from a previous experiment~\cite{bagheri} also appears to have excess strength, see Fig.~\ref{fig:RSF160}. Unfortunately, the strength function in the interesting region ($E_{\gamma} > 7$ MeV) suffers from 
poor statistics. However, this could be a hint that the same feature is present in this nucleus. 

\section{Summary and conclusions}  
The nuclei $^{163,164}\rm{Dy}$ have been investigated using the Oslo method. 
The radiative strength functions have been extracted, displaying the $M1$ pygmy resonance. This resonance has been studied in detail, and it is found that the measured widths are smaller than what has previously been measured in other Dy nuclei at OCL.  
However, the pygmy widths of $^{163,164}\rm{Dy}$ in the present work are still larger than what has been measured for $^{163}\rm{Dy}$ by means of the TSC method. When comparing 
the total integrated strength $B(M1\uparrow)$ of the $M1$ pygmy resonance, 
 the results for all the nuclei agree within the uncertainties.

For $^{164}\rm{Dy}$, we have observed an excess of strength for $E_{\gamma}~>~5$~MeV compared to model calculations; 
similar features can also be seen in $^{160}\rm{Dy}$. The enhanced strength might be due to neutron skin oscillations. If that is the case, this is the 
first time both the scissors mode and the neutron-skin oscillation mode is seen in one and the same nucleus.

\acknowledgments 
The authors would like to thank E.~A.~Olsen and J.~Wikne for excellent experimental conditions. Financial support from the Research Council of Norway (NFR) is gratefully acknowledged.

\end{document}